Identification of the major cause of endemically poor mobilities in SiC/SiO$_2$ structures

Xiao Shen[1] and Sokrates T. Pantelides[1,2]

[1] Department of Physics and Astronomy, Vanderbilt University, Nashville, TN 37235, USA

[2] Oak Ridge National Laboratory, Oak Ridge, TN 37831, USA

ABSTRACT

Materials with good carrier mobilities are desired for device applications, but in real devices the mobilities are usually limited by the presence of interfaces and contacts. Mobility degradation at semiconductor-dielectric interfaces is generally attributed to defects at the interface or inside the dielectric, as is the case in Si/SiO$_2$ structures, where processing does not introduce detrimental defects in the semiconductor. In the case of SiC/SiO$_2$ structures, a decade of research focused on reducing or passivating interface and oxide defects, but the low mobilities have persisted. By invoking theoretical results and available experimental evidence, we show that thermal oxidation generates carbon di-interstitial defects inside the semiconductor substrate and that they are a major cause of the poor mobility in SiC/SiO$_2$ structures.

--

Materials that demonstrate high carrier mobilities are broadly and actively studied for their novel properties and potential applications. Latest examples include graphene[1], epitaxial SrTiO$_3$ film[2], and GaAs quantum wells[3]. For practical applications, the relevant mobility is usually affected by the presence of interfaces. In particular, Si is the semiconductor of choice for microelectronics largely because thermal oxidation naturally leads to abrupt[4] Si/SiO$_2$ interfaces with benign defects, namely Si dangling bonds that are easily passivated by H. Virtually every other semiconductor is handicapped by the lack of a dielectric with good interfacial properties for the fabrication of metal-oxide-semiconductor field-effect transistors (MOSFETs) with large carrier mobilities. Silicon devices, however, are not very suitable for high-power, high-field or high-temperature applications because of the small energy gap and low thermal conductivity. Silicon carbide, especially the 4H polytype, has all the right features for such applications, including the unique feature that its native oxide is also SiO$_2$. Unlike the Si case, however, carrier mobilities at the SiC/SiO$_2$ interfaces are much smaller than their values in pure bulk SiC. More specifically, in Si/SiO$_2$ structures mobilities after H anneal are typically 50% of their bulk value whereas mobilities in thermal SiC/SiO$_2$ structures, after post-oxidation treatments, are at best 10-15% of the value in pure bulk 4H-SiC[5,6].

A key difference between oxidation of Si and SiC, both of which produce SiO$_2$, is that in the SiC case large amounts of C atoms are released[7,8]. It is generally believed that C atoms leave as CO or CO$_2$ through the growing oxide[9]. Residual C atoms in the SiC-SiO$_2$ interface region have been invoked as the primary origin of interfacial defects that limit electron mobilities[8-13]. During the 1990's oxidation of SiC followed by post-oxidation anneal in oxygen produced structures with electron mobilities of order 1-10 cm$^2$/V-s, compared with bulk values of ~800 cm$^2$/V-s in pure bulk 4H-SiC. In the last ten years, extensive experimental investigations led to significant improvements through post-oxidation annealing (POA) in NO[14,15] or H[10]. Even better mobilities were obtained recently by incorporating Na or P either during oxidation or by POA[5,6]. Still, the best mobilities obtained by oxidation of SiC are only 10-15% of the bulk value.

In a 2008 paper reviewing critical issues in SiC for power devices[16], Agarwal and Haney put forward a "conjecture" that defects in the SiC carbide that may be introduced during oxidation, high-temperature anneals, or ion implantation may contribute significantly to mobility degradation, but no particular defects were singled out. In 2008 and 2009, Zheleva *et al.* reported the presence of a "transition layer" in the SiC immediately adjacent to the SiC/SiO$_2$ interface with excess C concentrations as high as 20%, and found a dependence of the electron mobility on the thickness of the transition layer[17,18]. Subsequent theoretical simulations, however, found that such amount of excess carbon is likely to result in amorphization and C segregation[19].

There are good reasons, however, to expect non-negligible concentrations of C atoms to enter the SiC substrate as interstitials because the calculated migration barrier for C interstitials in SiC is only 0.5 eV[20]. Evidence in support of this hypothesis is provided by the recent experimental data of Hiyoshi and Kimoto[21]. Prior to oxidation, they used deep-level transient spectroscopy (DLTS) and detected the so-called $Z_{1/2}$ and $EH_{5/6}$ deep centers, which are generally believed to be the C vacancy or a carbon-vacancy-related defect[22]. After oxidation and removal of the oxide, they find that the concentration of these defects is substantially reduced, while another defect appears and grows to a significant concentration. They identified that defect as the so-called HK0 defect that had been detected in earlier work[23] and suggested that the HK0 defect is related to C or Si interstitials.

We note for comparison that during Si oxidation, Si interstitials ($Si_i$) are known to be emitted into both the Si substrate and the growing SiO$_2$ layer[24]. Excess Si atoms in Si are generally benign, as they diffuse with no energy barrier in the presence of electron-hole excitations[25] and the binding energy of di-interstitials $(Si_i)_2$ is relatively small, of order 1.8 eV[26], so that their formation at oxidation temperatures (850~1100°C) is unlikely. Carbon di-interstitials $(C_i)_2$ in SiC, on the other hand have a much larger binding energy, 5-6 eV[27,28], because of the formation of a strong double C bond[28]. A typical structure of $(C_i)_2$ is shown in Fig. 1. We recently performed finite-temperature simulations of a dilute C concentration in SiC and found that di-interstitial formation is rapid and barrierless[19]. We propose, therefore, that considerable concentrations of *immobile* C di-interstitials form in the SiC substrate during SiC oxidation and that in fact the HK0 center is the C di-interstitial. In the remainder of this paper, we examine in detail the properties of C di-interstitials and invoke both theoretical results and available experimental data to make the case that C di-interstitials is a major cause of mobility degradation in thermal SiC/SiO$_2$ structures.

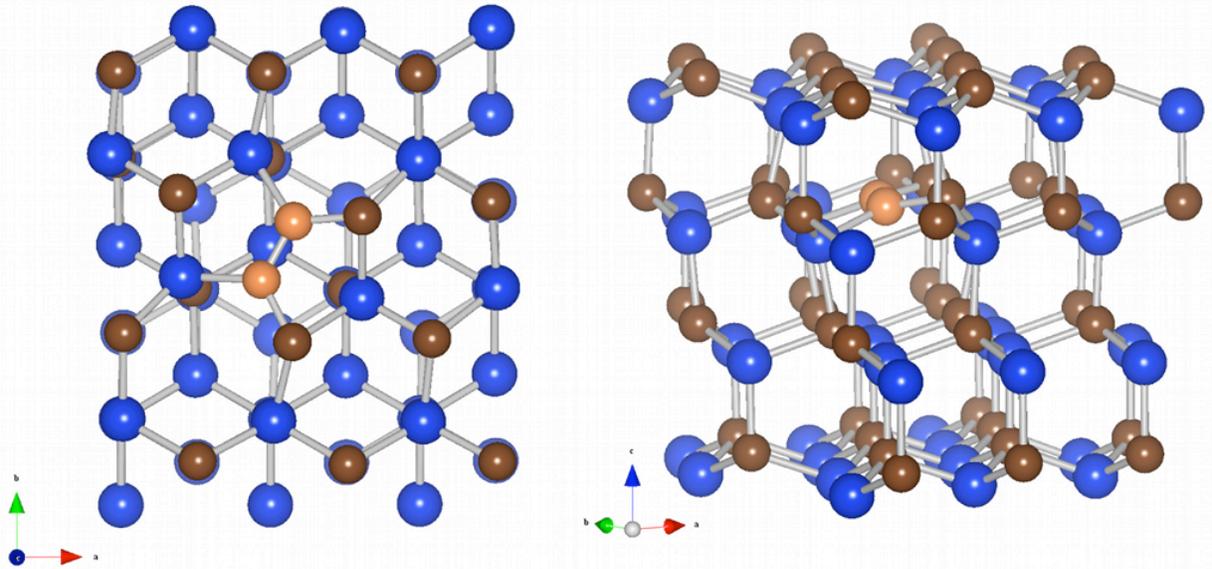

Figure 1. A typical structure of carbon di-interstitial cluster $(C_i)_2$ in bulk 4H-SiC. Left: Top view. Right: Side view. Silicon atoms are in blue, lattice carbon atoms are in brown, and interstitial carbon atoms are in copper. The planar bonding structure of $(C_i)_2$ shows that the two interstitial carbon atoms are sp2-hybridized, and are connected by a C=C double bond.

Recently, Devynck *et al.* calculated the energy levels for one $(C_i)_2$ defect using hybrid density functional, and found that it has four energy levels in the band gap: a (--/-) level at $E_c$-0.04 eV, a (-/0) level at $E_c$-0.23 eV, a (+/0) level at $E_v$+0.71 eV, and a (++/+) level at $E_v$+0.49 eV[29]. We note that the defect is negatively charged in n-type material and positively charged in p-type material, whereby it always limits carrier mobilities by Coulomb scattering.

Now let us examine the experimental reports of substrate traps. We already noted that data by Hiyoshi and Kimoto support the notion that C atoms are emitted into the SiC substrate and that, after annihilating possible C vacancies, lead to the formation of the HK0 defect. This defect is characterized by a level at $E_v$+0.78 eV, which matches well the theoretical (+/0) level of the C di-interstitial at $E_v$+0.71 eV. However, the (++/+) transition level, predicted to be at $E_v$+0.49 eV[29], is not observed[21], probably because it may be somewhat lower than the theoretical prediction and the temperature in the DLTS experiment was not low enough to observe it. It is known, on the other hand, that the HK0 center anneals in Ar at 1400 °C[30], which is consistent with the large (5~6 eV) binding energy of $(C_i)_2$[27,28]. The annealing behavior of the HK0 center also rules out the di-carbon antisite $(C_2)_{Si}$, since it is known to have smaller binding energy (~4 eV) and anneals out at lower temperature (1000 °C)[31].

Besides thermal oxidation, the HK0 center can also be created by reactive ion etching (RIE) followed by annealing[30], ion-implatation followed by annealing[32], and electron irradiation followed by annealing[23]. These processes create C interstitials inside SiC or at the surface. Furthermore, in the electron irradiation experiment, HK0 defects are created when the electron energy is above the carbon displacement threshold but below the Si displacement threshold[23]. This result rules out a possible role of Si interstitials and gives additional support for the notion that the HK0 center is a carbon interstitial cluster.

It is noteworthy that, while Al and Ne implantations followed by annealing create HK0 centers in SiC, N implantation does not[32]. Identification of the HK0 defect as $(C_i)_2$ provides an explanation of this observation: Prior theoretical study proposed that nitrogen interstitial $N_i$ can passivate $(C_i)_2$ defect at the interface[10]. The analysis of Ref. 10 was based on equilibrium configurations. Here we carried out quantum molecular dynamical (QMD) simulations to examine the non-equilibrium evolution of the system containing of $N_i$ and the $C_i$. The net conclusion is that $N_i$ interacts with both $C_i$ and $(C_i)_2$, thwarting the formation of stable $(C_i)_2$. In particular, $N_i$ binds to $(C_i)_2$ but then the complex releases a $C_i$ (Fig. 2). Thus, either $N_i$ attacks any HK0 that may have formed or simply captures the diffusive $C_i$ interstitials and pre-empts HK0 formation.

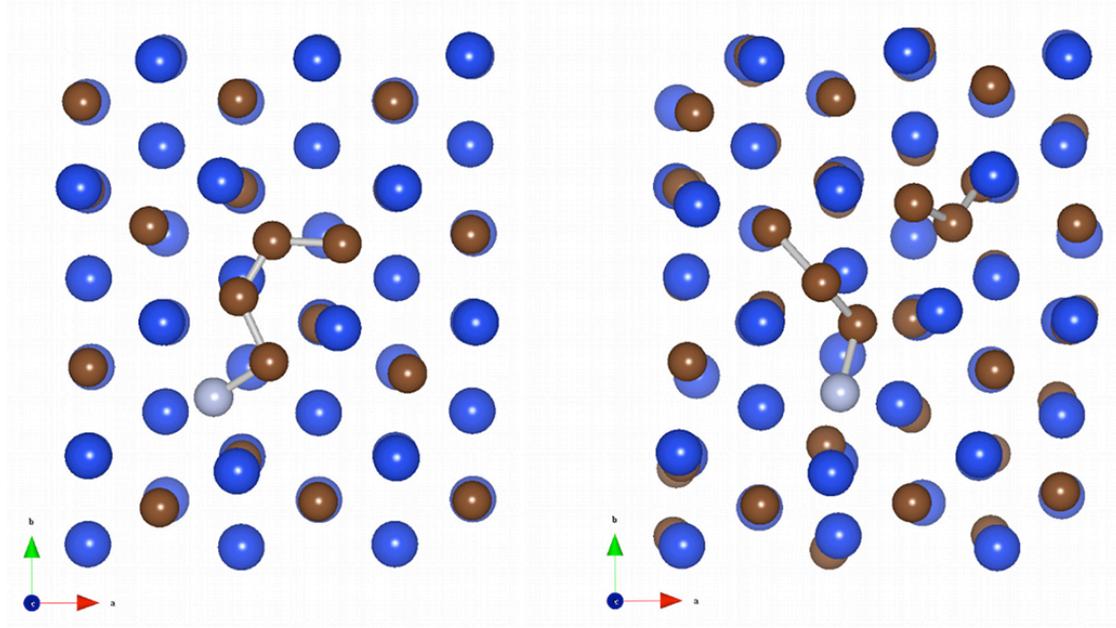

Figure 2. A quantum molecular dynamical (QMD) simulation showing a nitrogen interstitial $N_i$ attacking a carbon di-interstitial cluster $(C_i)_2$ in bulk 4H-SiC. Left: Starting position with $N_i$ bound to $(C_i)_2$ forming $N_i(C_i)_2$ complex. (Only short carbon-carbon and carbon-nitrogen bonds are shown.) Right: The complex breaks and one carbon atom has been emitted as a single interstitial $C_i$.

Tadjer *et al.* have reported a hole trap located in the SiC epilayer in 4H-SiC MOS capacitors and transistors[33]. The energy level of the hole trap is found to overlap with the Al acceptor level in SiC, which is around $E_v+0.22$ eV[34]. This hole trap may correspond to the (++/+) transition level of $(C_i)_2$. If this assignment is correct, then the absence of the (++/+) level in the data of Ref. 21 can be naturally explained, because the temperature range of the DLTS measurement reported in Ref. 21 is likely not low enough to observe this defect level.

Mooney and co-workers, using C-V and constant-capacitance DLTS (CCDLTS) measurements distinguished the traps in n-type 4H- and 6H-SiC MOS capacitors into oxide traps, real interface traps, and SiC substrate traps[35]. In NO annealed samples, the area densities of SiC substrate traps are found to be in the order of $10^{12}$ cm$^{-2}$, which is comparable to the densities of remaining oxide and interface traps. The substrate traps in 4H-SiC are found to have energy levels within 0.1 eV below the 4H-SiC conduction band. We propose that these traps correspond to the (-/0) and/or (--

/-) transition level of $(C_i)_2$ defect. Furthermore, unlike oxide traps and interface traps, whose densities decrease during NO annealing, the substrate trap densities increase during NO anneals. Although the experimental uncertainties are not available for those numbers, we suspect that the increase of substrate traps during NO anneal may be real. It is known that NO annealing dissolves carbon clusters at the SiC/SiO$_2$ interface[36]. This increase of substrate traps during NO anneal may be the result of dissolved carbon diffusing into the SiC and forming more $(C_i)_2$.

Dautrich, Lenahan and Lelis, using spin-dependent recombination (SDR) method, observed high concentration of deep level defects in 4H-SiC/SiO$_2$ MOSFETs below the SiC/SiO$_2$ interface into the SiC bulk[37]. The *g*-value for the observed defects is 2.0023, close to the *g*-value of a C dangling bond observed in disordered SiC (*g*=2.0027), suggesting that the electron wave functions are localized on C atoms, which can be either sp$^3$ bonded carbon or sp$^2$ bonded carbon[38]. The former is consistent with the tentative assignment of a Si vacancy given by the authors of Ref. 36, the latter is consistent with our hypothesis of $(C_i)_2$ defect. Further studies are required to clarify the nature of the observed C dangling bonds.

The CCDLTS data reported in Ref. 35 show that the area density of substrate traps is about $10^{12}$ cm$^{-2}$/eV. The assignment of these traps to $(C_i)_2$ clusters is consistent with the total interface density profile deduced from hi-lo C-V measurements[11,39]. The high concentration of total interface trap density near conduction band measured in n-type MOS capacitors may contain contributions from the (-/0) and (--/-) levels of $(C_i)_2$ clusters in the substrate. The high density of $D_{IT}$ in the lower part of the band gap (0.5 ~1.0 eV above valence band) measured in p-type MOS capacitors may come from the (0/+) level, and high density of $D_{IT}$ near valance band may contain contribution from the (++/+) level of $(C_i)_2$ clusters.

In principle, atomic H should passivate the $(C_i)_2$ defect by attacking the C=C double bond and forming C-H bonds. However, post-oxidation annealing with atomic hydrogen has no effect on all the three groups of $D_{IT}$ mentioned from $(C_i)_2$[10,39]. This result can be explained as follows: In SiC, the stable form of hydrogen is H$^+$ in p-type sample and H$_2$ in intrinsic and n-type samples, and only H$^+$ has a sufficiently low diffusion barrier[40]. Therefore, in n-type MOS capacitors, due to the low diffusivity of H$_2$, H cannot be incorporated into the substrate to passivate the $(C_i)_2$ defects; in p-type MOS capacitors, H can be incorporated as H$^+$, but the $(C_i)_2$ defects are also positively charged, whereby passivation is ineffective.

Okamoto *et al.*[6] found that the post-oxidation annealing (POA) with POCl$_3$ significantly decreases the $D_{IT}$ near the conduction band to the order of $10^{11}$ cm$^{-2}$/eV, which is much more effective than POA with NO. The difference between the effect of P and N annealing is still inconclusive. Nevertheless, the $(C_i)_2$ substrate trap picture provides a possible explanation. The present calculations show that, like N, P also can attack the $(C_i)_2$ defect and therefore reduce its concentration. However, while the diffusion of N into SiC is difficult[41], it has been shown that P can be easily incorporated into SiC. Diffusion of P from the surface into bulk 3C-SiC has been reported by Tin *et al.*[42]. Therefore, the more effective reduction of $D_{IT}$ may be a result of the more effective incorporation of passivators into the substrate.

Besides post-oxidation annealing techniques, recent experimental results reported by Ciobanu *et al.*[43], Poggi *et al.*[44], and Dhar *et al.*[45] show that pre-oxidation N implantation reduces $D_{IT}$ near the conduction band of SiC to the order of $10^{11}$ cm$^{-2}$/eV. We note that pre-oxidation N implantation is more effective in-reducing $D_{IT}$ than POA with N, which can be naturally explained if $(C_i)_2$ defects exist in the substrate. While POA with N is effective for passivating $(C_i)_2$ and other carbon clusters at the interface or in the oxide, it has little effect on the substrate defects because it is difficult to incorporate N from the surface. Meanwhile, implantation introduced plenty of N in the substrate, which can attack $(C_i)_2$ clusters and/or prevent their formation by capturing the injected $C_i$. We note that the latter process is more likely, since the oxidation temperature is not high enough to activate N diffusion.

The present proposal of $(C_i)_2$ in SiC as a major cause of low channel mobility suggests that deposited oxides, which avoid the emission of $C_i$, should yield higher mobility. However, SiC MOS structures with directly deposited $SiO_2$ may suffer from a poor interface and oxide quality, and post-deposition anneal often introduces additional oxidation of SiC[46]. In 2008, Hatayama *et al.* reported a remarkably high mobility that corresponds to ~35% of bulk value in SiC MOSFETs with deposited $Al_2O_3$ gate oxide[47]. The devices were fabricated by forming an ultrathin (~1 nm) layer of thermal $SiO_2$ and then depositing a ~70nm layer of $Al_2O_3$. The very thin thermal $SiO_2$ may have ensured a good interface with SiC while generated very little $C_i$. It is also reported that the channel mobility decreases rapidly when the thickness of thermal oxide increase to about 2 nm. This may due to the formation of a substantial concentration of $(C_i)_2$ in the substrate, since more $C_i$ are generated by oxidation.

In summary, we propose that the carbon di-interstitial cluster $(C_i)_2$ in SiC, generated by pairing of carbon interstitials during oxidation, is a major cause of the low channel mobility of thermlly grown SiC MOSFETs. It is an example of "channel traps", whose density, $D_{CT}$, should be minimized to improve the device performance. This defect can account for the SiC substrate traps reported in four independent experiments. Latest post-oxidation annealing and pre-oxidation implantation experiments are consistent with this picture. A recent report of very high mobility SiC MOSFETs with ultrathin thermal $SiO_2$ and deposited $Al_2O_3$ gate oxide also supports the present proposal.

Methods

We carried out finite-temperature QMD simulations using density functional theory. We employ the Perdew-Burke-Ernzerhof (PBE)[48] version of the generalized-gradient approximation (GGA) exchange-correlation functional for most calculations. We use projector augmented wave (PAW) potentials[49] and plane wave basis as implemented in the Vienna Ab-initio Simulation Package (VASP code)[50]. The calculations are done in a 3×4×1 96 atoms supercell of 4H-SiC. Static calculations with plane-wave cutoff of 400 eV and 2×2×2 Brillouin zone sampling are carried out to relax the structures. For QMD simulations, the plane-wave cutoff is 300 eV, and the Brillouin zone sampling is done with Γ point only. The time step for QMD simulations is set to 0.5 femtosecond. The simulation shown in Fig. 2 consists of heating the system to 2200 °C in 2.2 ps and performing constant temperature QMD for 3.9 ps.

Acknowledgement

We thank A. F. Basile, P. M. Mooney, Y. S. Puzyrev, B. R. Tuttle, and J. R. Williams for helpful discussions. This work is supported by NSF GOALI grant DMR-0907385 and by the McMinn Endowment at Vanderbilt University. Computational resources are provided by the National Science Foundation through TeraGrid resources under grant number TG-DMR100022.